# Electrocatalytic properties of manganese and cobalt polyporphine films toward oxygen reduction reaction

Dmitry V. Konev[1,2,*], Olga I. Istakova[1,2], Beata Dembinska[3], Magdalena Skunik-Nuckowska[3], Charles H. Devillers[4], Olivier Heintz[5], Pawel J. Kulesza[3,*], Mikhail A. Vorotyntsev[1,2,4,6,*]

[1] *D. I. Mendeleev Russian University of Chemical Technology University, Moscow, Russia*
[2] *Institute for Problems of Chemical Physics, Russian Academy of Sciences, Chernogolovka, Russia*
[3] *Faculty of Chemistry, University of Warsaw, Warsaw, Poland*
[4] *ICMUB, UMR 6302 CNRS-Université de Bourgogne-Franche-Comté, Dijon, France*
[5] *ICB, UMR 6303 CNRS-Université de Bourgogne-Franche-Comté, Dijon, France*
[6] *M. V. Lomonosov Moscow State University, Moscow, Russia*

**ABSTRACT**

Novel member of polymetalloporphines, namely manganese polymetalloporphine of type I (pMnP-I) obtained by ion exchange from magnesium polyporphine of type I (pMgP-I) is reported for the first time and compared to its cobalt analogue (pCoP-I). Both polymer films have been obtained via two-step procedure: demetalation of the pMgP-I electrode film via its exposure to trifluoroacetic acid solution, resulting in formation of the metal-free polyporphine of type I (pH$_2$P-I) followed by electrochemically induced incorporation of Co or Mn ions from the acetonitrile solution of cobalt and manganese perchlorates. A further oxidative transformation of pCoP-I, pMnP-I polymer films has led to the corresponding polyporphines of type II, pCoP-II and pMnP-II, possessing such unique features as condensed polymer structure with a very high density of active sites and high electronic conductivity within a very broad potential range including the one corresponding to the neutral (uncharged) state of the polymer matrix. Both polymers of type II also exhibit interesting electrocatalytic activity toward oxygen electroreduction in aqueous neutral (pH 6.7) and alkaline (pH 13) media which was evaluated under cyclic voltammetric and steady-state conditions. The results demonstrate that the efficiency (regardless of the electrolyte) of both polymetalloporphines is comparable to bare platinum electrode. The effect of annealing of polymer-modified electrodes on their catalytic properties has also been considered.

* Corresponding authors: dkfrvzh@gmail.com (D.V. Konev); pkulesza@chem.uw.edu.pl (P.J. Kulesza); mivo2010@yandex.com (M.A. Vorotyntsev)

**Introduction**

Various optical and electrochemical features as well as a high chemical and thermal stability make porphyrins promising materials for electronics, optoelectronics and photonics. These properties can be easily adapted to specific applications by introducing new metal ions in the center of the porphyrin ring or by chemical chain functionalization [1–3]. Due to the presence of active catalytic $MN_4$ centers (where M is $Co^{2+}$, $Mn^{2+}$, $Fe^{2+}$, etc. and $N_4$ is a porphyrin heterocycle), porphyrin based compounds found applications in various fields of electrocatalysis, chemical and biological sensing and molecular photovoltaic devices [4–10].

Electrochemical properties of metal porphines, i.e. the simplest, unsubstituted metalloporphyrins were for a long time barely known, mainly due to the lack of the commercial availability and multi-stage low-yield laboratory syntheses. A novel, three-step methodology of a relatively high yield synthesis of magnesium porphine (MgP), proposed by Lindsey et al. [11], significantly increased the interest in porphine and porphine-based materials. Thanks to this progress, it became possible to study both the behavior of these substances in the solution [12] and the processes resulting in formation of metalloporphine-based polymer films at electrode substrates, by potentiostatic or potentiodynamic electrooxidation of porphine monomers [13]. Such immobilization of porphines on solid electrode surfaces is of especial importance, in particular in a field of electrochemical sensing and electrocatalysis. An original approach based on a combination of electrochemical and spectroscopic methods allowed us to establish the molecular structure of such polymer films with Mg as a central metal ion. It has been demonstrated that during formation of the oxidation product, namely magnesium polyporphine of type I (pMgP-I) under a very low oxidation potential, each porphine ring forms ca. 2.1-2.2 bonds with adjacent monomer units. It implies that the polymer structure consists of chains of single *meso-meso* bonded monomeric porphine macrocycles, with a relatively small number of intermolecular bonds [14]. Comparative diagnostic studies of magnesium 5,15-di(p-methoxyphenyl)porphine and unsubstituted MgP polymerized under identical conditions have shown that contrary to the 5,15-substituted polyporphine, characterized by the linear chain structure, the structure of pMgP-I can be linear but also can contain zigzag and even cruciform fragments [15].

It has also been found that further oxidation of polyporphines of type I leads to formation of additional bonds between neighboring porphine rings, presumably in *beta-beta* positions. Such condensed tape structures (polyporphine of type II, pMgP- II) possess high electronic conductivity and redox activity within a very wide (over 3 V) potential range in organic media [16]. Furthermore, as previously demonstrated [17], the efficiency of the polymerization process can be increased by introducing a proton-acceptor additive into the electropolymerization medium, such as 2,6-dimethylpyridine (lutidine) which enhances the rate of film formation by the factor of two while it is not included inside the polymer film.

As aforementioned, owing to their unique properties polyporphines containing electrocatalytically active metal sites may serve as powerful electrocatalysts and redox mediators supporting various oxidation/reduction processes. Therefore, our recent research has been focused on developing procedures of ion exchange replacement in pMgP-I polymeric coatings as a means to incorporate therein transition metal ions, with conservation of the molecular structure of the starting polymer. It has been achieved by demetalation of the pMgP-I polymer with trifluoroacetic acid of hydrochloric acid, leading to the formation of "free-base" polyporphine of type I, $pH_2P$-I. Subsequent treatment of demetalated polymer films by saturated Zn or Co ion containing solutions in relevant temperature intervals resulted in relatively fast (within 1 h) formation of the corresponding polymetalloporphines, pZnP-I or pCoP-I [18,19]. It is noteworthy that ion-exchange procedure does not change either the initial polymer morphology or its molecular structure, as well as the polymer electroactivity is preserved within a broad potential range. In order to optimize the preparation methodology, a novel synthetic approach has been developed recently, based on electrochemical (potentiodynamic) treatment of the metal-free polyporphine coating, $pH_2P$-I, in contact with a solution containing the corresponding metal ion [20,21]. Such a synthetic route leads to much faster film preparation, besides with the use of dilute solutions and ambient temperature.

$MN_4$-type macrocyclic metal complexes, in particular cobalt and iron porphyrins and phthalocyanines, are well known as oxygen reduction catalysts [22-28]. Compared to most other materials containing these complexes, polyporphines possess certain benefits such as very high density of active sites due to the absence of spacers (cross-linking groups) between monomer units, spatial matrices and side substituents. Moreover, the fact that porphine centers are situated inside the main chain results in a high degree of conjugation through which the electrons can be transferred from the electrode to the substrate and back. Our preliminary studies have shown that indeed, pCoP films of both type I and type II exhibit electrocatalytic properties towards oxygen reduction reaction in alkaline media [29]. It is noteworthy that for the Co(II) polyporphine of type II, the $E_{1/2}$ potential of the oxygen reduction reaction was very close to the value characteristic of platinum electrode. Current densities were also higher than those for the Co polyporphine of type I which has been attributed to much higher electronic conductivity of these films within a wider potential range, allowing for efficient electron transport between the active sites in the polymer film and a substrate to take place.

Manganese cations also belong to catalytically active transition metal ions. Therefore, based on existing preparative methods and our preliminary studies, electrocatalytic films of cobalt and manganese polyporphines of type II (pCoP-II and pMnP-II) have been prepared in this study with the use of electrochemically induced ion exchange. Then, their electrochemical and electrocatalytic activities in the oxygen reduction reaction have been evaluated under cyclic voltammetric and steady-state conditions (polarization plots) in aqueous (alkaline and neutral) media. These electrocatalytic properties have been compared to that of the bare platinum electrode. Physicochemical properties of manganese polyporphine, reported for the first time, have been established using XPS and IR spectroscopies.

**Experimental**

*Materials and electrochemical equipment*

Magnesium porphine (MgP) monomer was synthesized according to Lindsey's procedure [11]. Acetonitrile (AN) of HPLC classification (Panreac) was dehydrated with molecular sieves (pore size 3Å) as drying agent. Background electrolyte, tetrabutylammonium hexafluorophosphate ($TBAPF_6$, Fluka), was stored in an oven at 80° C for 24 hours prior to use. 2,6-dimethylpyridine (lutidine, 99%) and trifluoroacetic acid (TFAA, >99%) were purchased from Sigma Aldrich and tetraethylammonium perchlorate ($TEAClO_4$, Carlo Erba, 98%) was from Carlo Erba.

All electrochemical measurements were performed using Elins PI50PRO3 electrochemical workstation in the conventional three-electrode glass cell equipped with glassy carbon or platinum disc working electrode (geometric area of 0.0078 $cm^2$) and large surface area platinum foil counter electrode (separated from the AN solution by porous glass membrane). For experiments in acetonitrile (AN), silver $Ag/Ag^+$(AN) electrode ($10^{-2}$ mol $dm^{-3}$ $AgNO_3$ + 0.1 mol $dm^{-3}$ $TBAPF_6$ in AN) separated by double frit filled in with background solution (0.1 mol $dm^{-3}$ $TBAPF_6$ in AN) was used as reference electrode. Its potential was 0.1 V more negative than the formal potential of the $Fc/Fc^+$ redox couple in the same solvent. For studies in aqueous electrolytes, Ag/AgCl/ sat. KCl (0.198 V *vs*. SHE) served as reference electrode.

*Deposition and characterization of polymetalloporphines (pMPs)*

Prior deposition and electrochemical characterization, all solutions were deaerated under vacuum and subsequently saturated with high-purity argon in the Schlenk system.

Starting films of magnesium polyporphine of type I (pMgP-I) were deposited *via* potentiostatic (0.35 V vs. $Ag/Ag^+$) electropolymerization of the MgP monomer from its 0.5 mmol $dm^{-3}$ solution in AN with the addition of 0.1 mol $dm^{-3}$ of $TBAPF_6$ and $1.5·10^{-3}$ mol $dm^{-3}$ of lutidine as proton-acceptor additive, according to the previously reported procedure [13]. These conditions correspond to formation of the type I structure formed by Mg(II) containing porphine cycles which are singly bonded in *meso-meso* positions. Thickness of the film was controlled by the charge passed during deposition where 5 mC $cm^{-2}$ corresponds to the thickness of ca. 50 nm.

To prepare films of cobalt and manganese polyporphines of type I (pCoP-I and pMnP-I), the magnesium polyporphine (pMgP-I)-coated electrodes were at first demetalated in AN solution of trifluoroacetic acid (1:50 v/v) according to the earlier developed procedure [18]. Incorporation of metal ions into the structure of demetalated polyporphine (pH$_2$P-I) films was achieved with the use of a recently proposed method of electrochemically induced ion exchange [18]. Briefly, for preparation of pCoP-I and pMnP-I films, metal-free polyporphine-modified electrode (pH$_2$P-I) was metalated in contact with 0.1 mol dm$^{-3}$ TBAPF$_6$ solution in AN containing 5·10$^{-4}$ mol dm$^{-3}$ metal salt: either Co(ClO$_4$)$_2$ or Mn(ClO$_4$)$_2$, by performing 30 voltammetric cycles in the potential range between -1.2 V (for Co) or -1.4 V (for Mn) and +0.4 V (starting potential of the procedure: 0 V) with scan rate of 0.1 V s$^{-1}$.

Oxidative transformation of pCoP-I and pMnP-I into polymetalloporphines of type II with condensed structure (pCoP-II and pMnP-II) was performed under galvanostatic conditions in AN solution of 0.1 mol dm$^{-3}$ TBAPF$_6$ until the potential reached 1 V. All steps of this procedure are summarized in Scheme 1.

Fourier-transform infrared spectra under attenuated total reflectance mode (FTIR-ATR) were measured by Bruker Vertex 70v vacuum Fourier IR spectrometer equipped with diamond optical element. Various polyporphine films, pMgP-I (deposition charge: 5mC cm$^{-2}$), pH$_2$P-I and pMP-I (M = Co or Mn) were prepared on 25 μm thick platinum foil following the above procedure (Scheme 1). Then, their spectra were measured with resolution of 4 cm$^{-1}$ and averaged over 50 scans.

Samples for XPS measurements were prepared analogously. XPS spectra were recorded with the use of SIA100 device (Cameca Riber apparatus) with non-monochromatized Al Kα source (1486.6 eV).

*Electrocatalytic properties*

Electrocatalytic properties of cobalt and manganese polyporphines films toward oxygen reduction reaction were studied in 0.1 mol dm$^{-3}$ tetraethylammonium perchlorate (TEAClO$_4$, pH=6.7) and 0.1 mol dm$^{-3}$ NaOH (pH=13) solutions under voltammetric regime as well as under steady-state conditions: potential step of 0.05 V, potential hold time of 20-50 s, i.e. until the current reached plateau.

To illustrate the influence of the annealing treatment on the activity and stability of pCoP-II and pMnP-II coatings, the metalation of metal-free polyporphine of type I, pH$_2$P-I (corresponding to step 3 in Scheme 1) was performed *via* chemical method [18,19,29,30], i.e. by holding the pH$_2$P-I-modified electrode in dimethylformamide (DMF) solution of 5·10$^{-2}$ mol dm$^{-3}$ of CoCl$_2$ or Mn(OAc)$_2$ for 1 h at 130°C. Subsequent oxidative transformation of type I coatings into those of type II was performed galvanostatically, as already described. The samples were subsequently subjected to pyrolysis of the polymers by annealing at 800°C for 2 h. Glassy carbon rods (diameter: 2mm; length of immersed part: ca 3mm; geometrical surface area: ca 0.2 cm$^2$) were used as substrate. Their surface was roughened by means of abrasive paper (grain size of about 30 μm) to prevent possible peeling of the polymer layer during annealing process.

## Results and discussion

### *Fabrication and electrochemical identity of polymetalloporphines*

Introduction of manganese (II) and cobalt (II) ions into the $pH_2P$-I polyporphine film deposited on platinum electrode was conducted with the use of the electrochemically induced ion exchange [20,21]. Since electrochemical and physicochemical properties of cobalt polyporphine have already been presented in details, here only the formation of its manganese analogue is shown as a representative example. Cyclic voltammetry response in the course of the electrochemically induced ion exchange (resulting in formation of pMnP-I) achieved upon exposure of the metal-free $pH_2P$-I polyporphine deposit to $Mn(ClO_4)_2$ containing solution is presented in Fig. 1. Almost complete disappearance of n-doping currents of $pH_2P$-I (in the potential range below -1.25 V) [18,20,21] and some increase in the electroactivity of the film at more positive potentials (increase of capacitive-type currents) can be seen during consecutive potential cycling. There is also a small shift and broadening of weak current responses (from -0.85 to -0.67 V for the negative potential sweep and from 0.03 to 0.1 V for the positive one), the existence of which is attributed to so-called "trapped charges" [18,20,21].

In order to confirm that manganese ions have successfully been incorporated into the polymer films, FTIR-ATR spectra of 50 nm-thick coatings of polyporphines of type I, namely, initial pMgP-I, metal-free $pH_2P$-I and pMnP-I deposited on the surface of platinum foil were recorded (Fig. 2). Three regions, A, B and C, have been distinguished to illustrate changes in the vibrational bands at ca. 1050, 1000 and 780-850 $cm^{-1}$, respectively. The bands appearing within regions A and C of the spectra are related to vibrations of different symmetries of external hydrogen atoms in the porphine plane and out of it, respectively. In conformity with expectations, demetalation process does not affect the vibrational bands in these regions (Figs. 2a and 2b) [18,31,32]. The most pronounced changes for demetalated polyporphine, $pH_2P$-I, can be seen in region B: splitting of the intensive signal at ca. 1000 $cm^{-1}$ for pMgP-I (Fig. 2a) into three bands of lower intensities shifted towards lower wavelengths for $pH_2P$-I (Fig. 2b). The signal is related to in-plane deformational vibrations of macrocycles with most intensive displacement of nitrogen and carbon atoms of pyrrole rings. The observed changes clearly indicate that demetalation has been accomplished, resulting in lowering the symmetry of porphine units upon replacing of Mg cation by two protons.

As can be seen in Fig. 2c, single band with the intensity comparable to pMgP-I (Fig. 2a) appears again in region B after passage of the remetalation process. It means that the porphine symmetry has been restored to its initial state, i.e. that the Mn cation has successfully been incorporated into the structure. However, the existence of a small signal at 970 $cm^{-1}$ (Fig. 2c), also observed for $pH_2P$-I at the same wavelength might imply that metalation process has not been fully completed. It was established earlier for pCoP-I [21] that the degree of metalation achieved via this procedure depends on the film thickness. According to Fig. 2c the same effect is also observed for pMnP-I.

XPS spectra of various polyporphines of type I, pMP-I (M = Mg, 2H, Co or Mn, Fig. 3), deposited on platinum foil also confirm the presence of Co or Mn ions inside the coating after the electrochemically induced ion exchange in contact with solutions of the corresponding salt. This conclusion is based on both the existence of characteristic peaks of Co or Mn (Fig. 3a) and on the shape of N 1s band (395-400 eV) as shown in Fig. 3b. As seen, a single peak for pMgP-I due to 4 equivalent nitrogen atoms of deprotonated pyrrole groups is split into doublet for $pH_2P$-I due to two non-equivalent groups (containing 2 deprotonated and 2 protonated pyrrole nitrogen atoms) at binding energies of 398 and 400 eV, respectively [33]. Incorporation of the metal ion (Co or Mn) into porphine units restores the equivalence of all 4 nitrogen atoms owing to identical bonds between the ion and each nitrogen atom.

Results for the atomic compositions (C, N, Mg, Co, Mn) of various polyporphines, on the basis of the XPS spectra, are given in Table 1. Calculated ratios of the atomic contents for C to N and N to M show that the degree of metalation, i.e. the fraction of Co or Mn containing porphine units, is close to the complete one as a result of the electrochemically induced ion exchange procedure. The difference in conclusions on this point on the basis of the IR and XPS spectra may be explained in terms of the practically complete metalation of the external surface layer of a film (responsible for the XPS signal) while the degree of the metal ion incorporation into porphine units in the depth of the film (affecting the IR signal in Fig. 2) may be lower. This conclusion is in conformity with the earlier observation [21] that the metalation degree is lower for thicker polyporphine films.

Comparative cyclic voltammetric characteristics of the pMnP-I and pCoP-I films (deposited onto platinum electrodes) recorded in 0.1 mmol $dm^{-3}$ $TBAPF_6$ + AN solution are shown in Fig. 4a. It is evident that in case of pCoP-I (black curve in Fig. 4a) two sets of redox peaks, located at around -1.25 V and 0.1 V, are well pronounced. They can be ascribed to Co(I)/Co(II) and Co(II)/Co(III) transitions, respectively [19]. It is also apparent that these peaks are separated by a large potential interval of low electroactivity of the polymeric film (low capacitive-type currents) where it exists in the neutral insulating state [13,19]. In the case of pMnP-I (red curve in Fig. 4a), there is no signal in the potential range (expected on the basis of the corresponding monomeric porphyrins) corresponding to the redox transition of the central Mn atom, Mn(II)/Mn(III) [27,34] but, in comparison to pCoP-I, the electroactivity of the film is higher in the whole intermediate potential range. The absence of the characteristic peak related to the redox transition of Mn ions presumably originates from the fact that it should take place in the potential range where the polymeric matrix has a low conductivity, thus the electron transfers between the electrode and Mn centers are practically inhibited.

In our previous reports we have shown that polyporphines of type II, pMP-II (M = $2H^+$, $Mg^{2+}$, $Zn^{2+}$, $Co^{2+}$), are characterized by a substantially higher electrical conductivity than polyporphines of type I, pMP-I [17,18]. As apparent from Fig. 4b, where cyclic voltammetric responses of the pMP-II films recorded in contact with background solution are presented, the range of the electroactivity (observed as increased capacitive-type currents) is broadened for both pCoP-II and pMnP-II polymetalloporphines in comparison to those of type I (Fig. 4a). For the pCoP-II film, well-pronounced peaks originating from Co(II)/Co(III) redox transitions (at around 0.1 V) are preserved while the signals corresponding to Co(I)/Co(II) couple are less distinct, compared to the response of the pCoP-I film (black lines in Figs 4a and b). Probably the Co(I)/Co(II) signals in pCoP-II are masked by increased capacitive currents of the polymer. As further seen for pMnP-II, highly reversible peaks corresponding to the Mn(III)/Mn(II) couple (located at around -0.5 V) appear while the corresponding signals were absent for the pMnP-I film (red curves in Figs 4a and b). The observation confirms that the electrical conductivity of polymetalloporphines of type I, pMP-I, considerably increases within a broad potential range upon their conversion into pMP-II, irrespective of the central metal ion as it has recently been proven for Mg, Zn and Co containing films [17,18] and for the first time is presented here for Mn. Fig. 4c illustrates the potential-time dependence for the pCoP-I and pMnP-I films in the course of their galvanostatic oxidative transformation into type II. Similarity of these curves as well as of the charges passed during the transformation for these films testify in favor of the identical mechanism of this process for all polyporphines which consists in formation of two *beta-beta* bonds between every couple of neighboring porphine units inside polymer chain. Further studies of electrocatalytic properties have been performed for the films of type II.

*Electrocatalytic activity of pMnP-II and pCoP-II toward oxygen reduction*

The electrocatalytic efficiency toward oxygen reduction of pMnP-II and pCoP-II films was examined in neutral and alkaline aqueous solutions of 0.1 mol dm$^{-3}$ TEAClO$_4$ (pH 6.7) and 0.1 mol dm$^{-3}$ NaOH (pH 13), respectively. A glassy carbon was used as a support for polymetalloporphine films since platinum is known as a highly efficient oxygen electroreduction catalyst so that the origin of the catalytic activity would be unclear. In contrast, the oxygen reduction at bare glassy carbon is rather poorly efficient and starts at much more negative potentials. Prior to experiments in oxygen-saturated solutions, the electrochemical properties of pMnP-II and pCoP-II polymer films were at first examined in deaerated aqueous electrolytes of different pH (Figs 5a,b). For pCoP-II, a pair of broad symmetrical peaks, appearing in the potential ranges from 0 to 0.4 V in neutral (black curve in Fig. 5a) and from -0.4 to 0 V in alkaline (black curve in Fig. 5b) solutions, most likely reflects redox transitions of cobalt cations. In the case of pMnP-II in neutral medium, the redox response which should be attributed to transitions of Mn cations, can be noticed as poorly developed waves at potentials below -0.1 V (red curve in Fig. 5a). On the other hand, the same couple (evident in the potential range between -0.8 and -0.3 V) is much better pronounced in alkaline media (red curve in Fig. 5b).

Figs 6 and 7 show conventional cyclic voltammograms as well as steady-state polarization curves for the oxygen electroreduction in neutral (a) and alkaline (b) solutions at pMnP-II and pCoP-II films deposited on glassy carbon electrodes (black and red curves, respectively). For comparative purposes, results of analogous experiments conducted for unmodified glassy carbon and platinum electrodes (having the same geometrical surface areas) are also shown in Figs 6 and 7 (green and blue curves, respectively). Comparison of the heights of characteristic peak currents and limiting currents with those obtained for bare platinum and glassy carbon electrodes allows to roughly predict the number of electrons participating in the electrocatalytic reaction, i.e. assigning it to 2- or 4-electron mechanism. As seen, the peak (Fig. 6) as well as the steady-state (Fig. 7) currents of the oxygen reduction for both polyporphine coated electrodes, pCoP-II and pMnP-II, are of the same amplitudes as those of bare platinum, while they are more than twice as high as the currents observed for bare glassy carbon electrode. This observation would suggest primarily the 4-electron reaction mechanism of the process (with formation of water as the main product). However, it is well established that the activity of macromolecular complexes depends on numerous factors such as central metal ion, ligand, substituents in the ring, support or the preparation methods of the modifying layer [35-39]. It has been proven for instance that, in contrast to most monomeric cobalt porphyrins (and other macrocyclic compounds), the simplest Co porphine (not polymerized) is able to reduce oxygen directly to water in acidic electrolyte (due to spontaneous dimerization of porphine units), but irreversible oxidation of the porphine ring leads to 2-electron reduction of $O_2$ catalyzed by the complex [35]. In other studies it was revealed that unstable monomeric Co(II) tetra-aminophthalocyanine, which participates in the 2-electron oxygen reduction, turns into stable catalyst capable of reducing oxygen to water after its polymerization [36]. Rather scare data for manganese macromolecular complexes imply the promotion of 2-electron reduction of oxygen in neutral and 4-electron reduction in basic solutions, while electropolymerized tetratolyl porphyrin drives the reaction to water in neutral media [37-39]. Here we can further go with your explanation suggesting rather 4-electron reduction mechanism, will it be ok?

The efficiency of the oxygen electroreduction was estimated based on both the peak potential of cyclic voltammetric curves (Fig. 6) and the half-wave potential of steady-state voltammograms (Fig. 7). The results for various electrode materials are summarized in Table 2. As seen, in both neutral and alkaline electrolytes the potentials values obtained for pMnP-II and pCoP-II catalytic films approach those characteristic of compact platinum which indicates the high electrocatalytic efficiency of these polymetalloporphines towards oxygen reduction. The most positive potentials of the process, that even slightly exceed that of compact platinum (at least under voltammetric conditions), are observed for pCoP-II (Fig. 6, Table 2).

It is noteworthy that the electrocatalytic activity of metal macrocyclic complexes presented here is practically identical in terms of the limiting current (0.3-0.4 mA cm$^{-2}$) compared to those for monomeric [23] or electropolymerized [27] metalloporphyrin-modified porous electrodes (containing graphene oxide frameworks or carbon nanotubes), in spite of the fact that the thickness of pMnP-II and pCoP-II was on the level of 50 nm, and thus the overall concentration of active sites per 1 cm$^2$ of the surface area was very low. Moreover, the characteristic potentials for polymetalloporphines of type II are more positive than the corresponding ones (for the same metal ion) for electropolymerized porphyrins [27].

What is more, in contrast to results concerning metalloporphyrins [23] or polymerized metalloporphyrins [27], the activity of our Mn-containing macrocyclic compound is close to the activity of its Co counterpart in terms of both the limiting current and the potential (merely 0.1 V negative shift). These observations may originate from the unique structure of polymetalloporphines of type II characterized by the highest possible density of active $MN_4$ sites and excellent electron conductivity, compared to conventional macromolecular complexes like porphyrins or phthalocyanines.

The most significant problem of $MN_4$-type catalytic centers with respect to oxygen reduction reaction is their poor stability. To estimate the stability of Co and Mn polyporphines, the pMP-II films were subjected to potential cycling in contact with aerated electrolyte solutions (data not shown). For both pMnP-II and pCoP-II-modified electrodes, the peak currents decrease and a slight displacement of peak potentials toward lower potential values were clearly observed after several voltammetric cycles, especially in neutral media. The phenomenon might imply a certain decrease in the number of catalytically active sites due to aggressive action of various peroxides, i.e. intermediate products of the oxygen reduction reaction. Furthermore, it should be kept in mind that the applied electrolyte, $TEAClO_4$, does not possess buffering properties, thus local pH changes may clearly occur. Even though its pH was 6.7, i.e. the content of protons in the bulk solution was very small, a progressive demetalation of polyporphine ring may proceed. Besides, due to the presence of the $H_2O_2$ intermediate, the Fenton-type reaction (generating radical oxygen species) may take place [40]. What important, both polymetalloporphines were more stable in alkaline media, probably since the demetalation of macromolecular complexes is avoided, which is a well-known feature of catalysts based on $MN_4$-type centers.

Conventional treatment for the stability enhancement of the transition metals coordination compounds is the annealing (typically at 800–900°C) of the complex, previously adsorbed or polymerized on the surface of carbon support [22,41-44], leading to their pyrolysis. As a result, metal-nitrogen-carbon type catalysts are produced in which $MN_x$ structures are present inside the carbon matrix. Fig. 8 shows comparison of voltammetric curves of oxygen electroreduction proceeding at the films of pMnP-II and pCoP-II recorded in contact with 0.1 mol dm$^{-3}$ alkaline electrolyte before and after annealing at 800 °C.

In these experiments, a glassy carbon rod with a rough surface was used as substrate in order to prevent possible peeling of the polymer layer during annealing process. This roughness explains an increase in the capacitive contribution to the observed currents in argon-saturated solutions.

It is evident from Fig. 8a that the thermal treatment of the polyporphine, pMnP-II, with its pyrolysis and carbonization, leads to almost complete loss of its activity towards oxygen reduction, compared to the non-heated sample, which confirms previous observations that Mn (in opposite to most macrocyclic metal complexes) may lose its $N_4$ structure [38]. In case of pCoP-II, as expected, the stability of the voltammetric responses in the oxygenated solutions in the course of the multi-cycle procedure (at least within a short timescale) has increased. However, the comparison of the currents of oxygen electroreduction shows that the electrocatalytic effect is also weakened after annealing of pCoP-II (Fig. 8b) although to a much less extent than that for its Mn analogue. As a whole, it can be concluded that for the newly obtained coatings of polymetalloporphines, the annealing process is not an efficient route to increased efficiency of the oxygen reaction process since it leads to the significant deterioration in electrocatalytic current densities as well as to the shift of the oxygen electroreduction potential towards more negative values.

**Conclusions**

A novel representative of the polyporphine family containing Mn(II) cations as central ions (both of type I, pMnP-I, and type II, pMnP-II) have been obtained for the first time via our procedure of the electrochemically induced ion exchange. It has been demonstrated that manganese and its cobalt counterpart polyporphines of type II (pMnP-II and pCoP-II) existing as 50 nm thin films on solid electrode substrates can be considered as promising platinum-free electrocatalysts for the oxygen reduction in neutral and especially (due to a higher stability) alkaline electrolytes. The measurements conducted under cyclic voltammetric and steady-state conditions have shown that regardless of the electrolyte pH, the pCoP-II polymer is more efficient catalyst toward oxygen reduction reaction and its performance approaches to that of compact platinum. However, it is of importance to note that (contrary to literature data concerning similar porphyrin complexes) the activity of manganese-containing polyporphine of type II, pMnP-II, is quite close to the performance of its Co counterpart. The high electrocatalytic activity of both electroactive polymers should be ascribed to the unique structure of polymetalloporphines of type II, i.e. the highest possible density of active $MN_4$ sites and a high electron conductivity within a very broad potential range.

**Acknowledgement**

Financial support of the Russian Foundation for Basic Research (grant 16-03-00916 A) is acknowledged. Partial support from the European Commission through the Graphene Flagship – Core 1 project [Grant number GA-696656] is also appreciated.

**Figures**

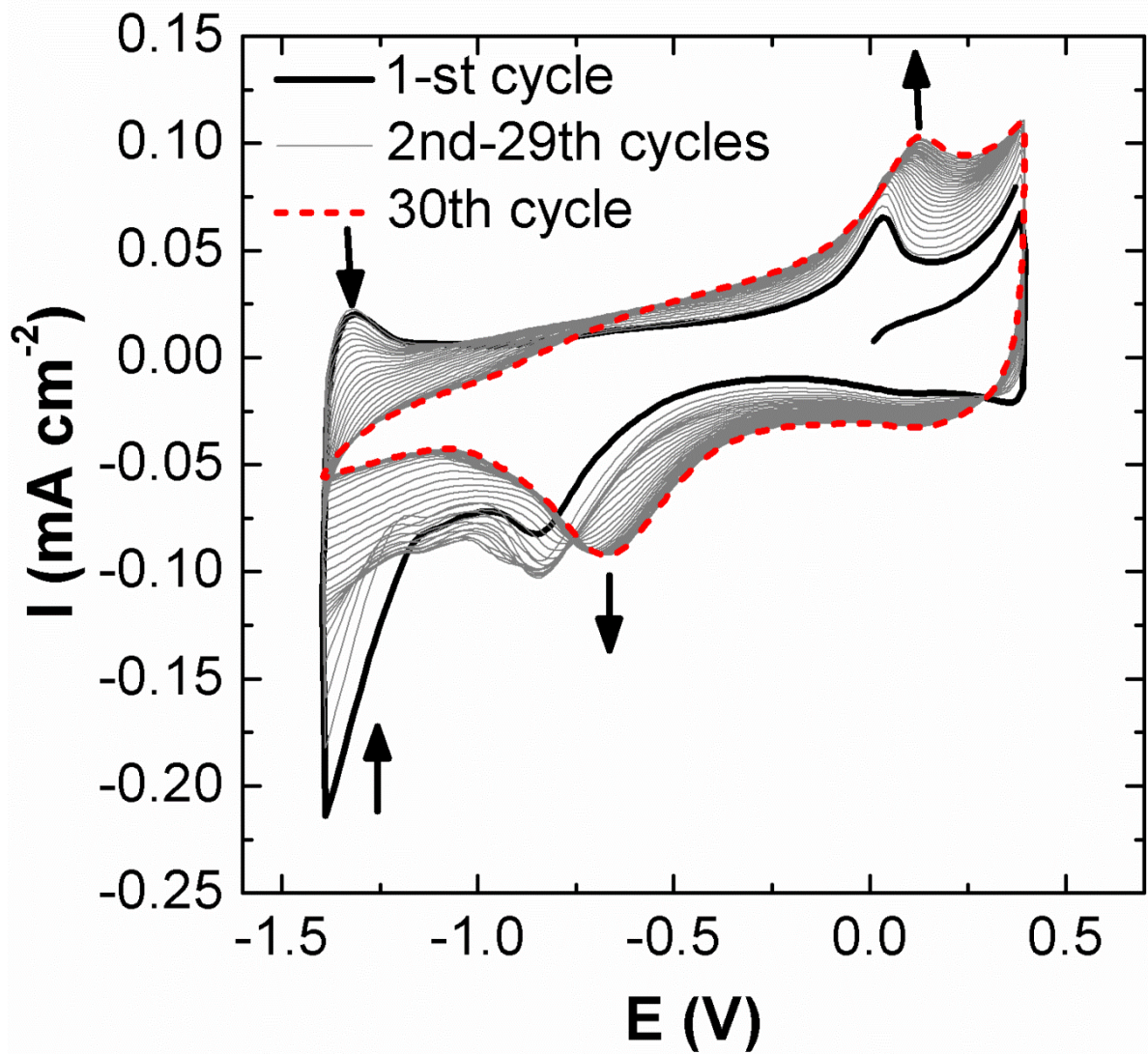

**Fig. 1.** Cyclic voltammetric responses of platinum electrode modified with pH$_2$P-I (deposition charge of initial pMgP-I was 5 mC cm$^{-2}$ which corresponds to the thickness of ca. 50 nm) in the course of electrochemically induced ion exchange in contact with 5·10$^{-4}$ mol dm$^{-3}$ Mn(ClO$_4$)$_2$ containing AN solution. Potential sweep rate: 0.1 V s$^{-1}$.

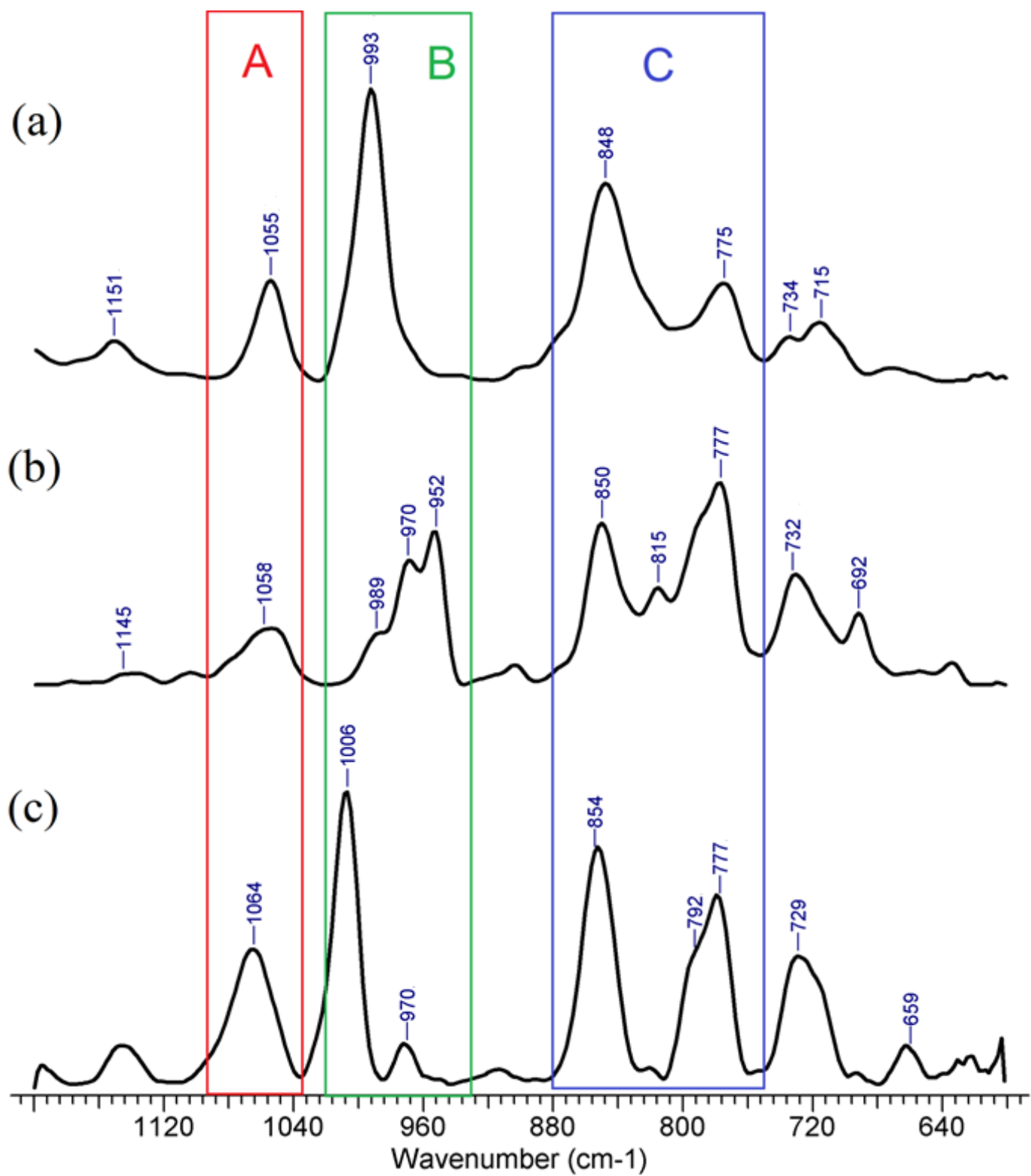

**Fig. 2.** FTIR ATR spectra of polyporphines of type I: (a) pMgP-I, (b) pH$_2$P-I, (c) pMnP-I deposited on platinum foil.

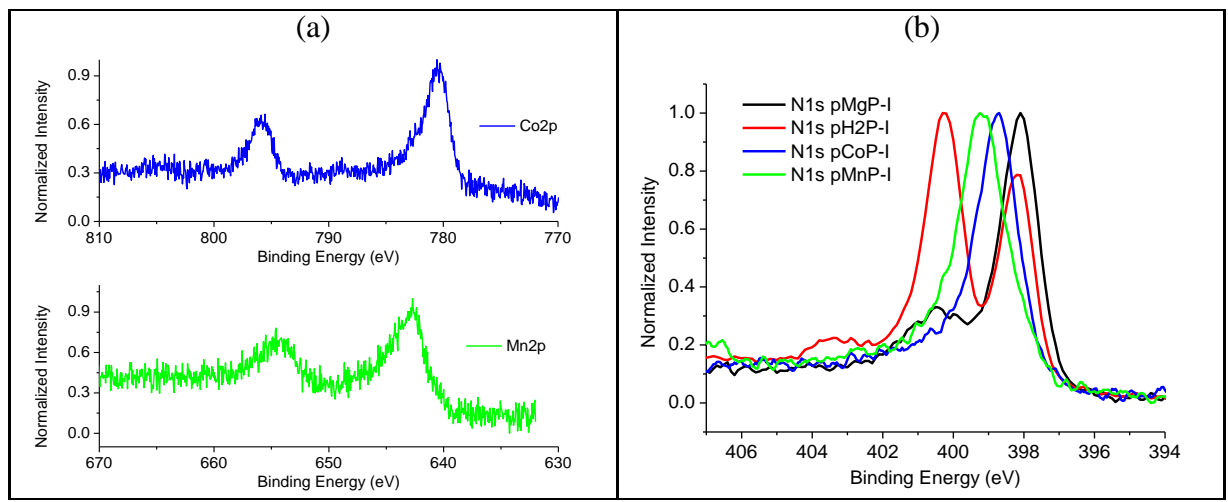

**Fig. 3.** Fragments of XPS spectra of pMP-I films deposited on Pt foil. Deposition charge of initial pMgP-I was 5 mC cm$^{-2}$ which corresponds to the thickness of ca. 50 nm. (a): bands M 2p (M = Mn or Co), (b):band N 1s

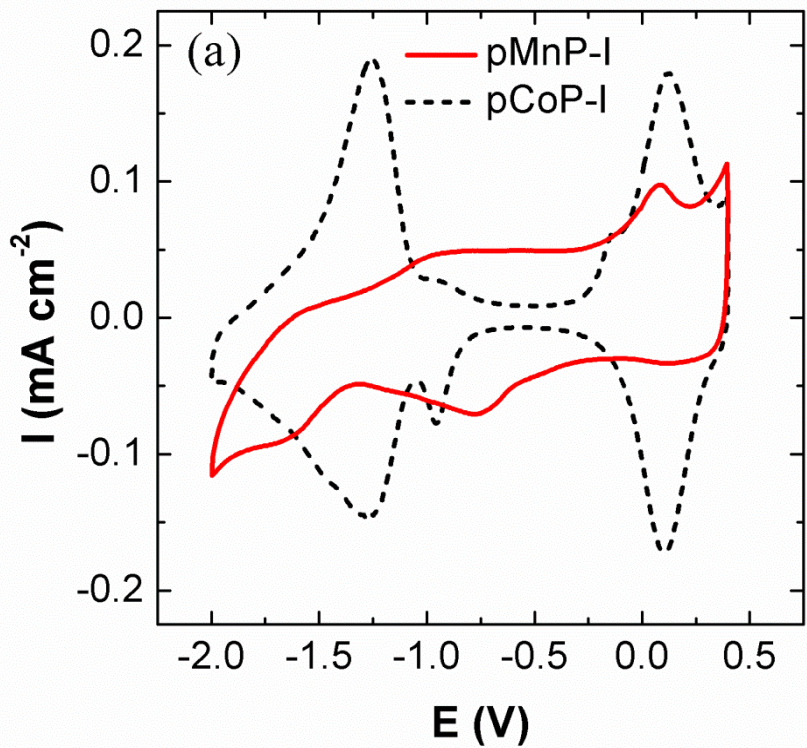

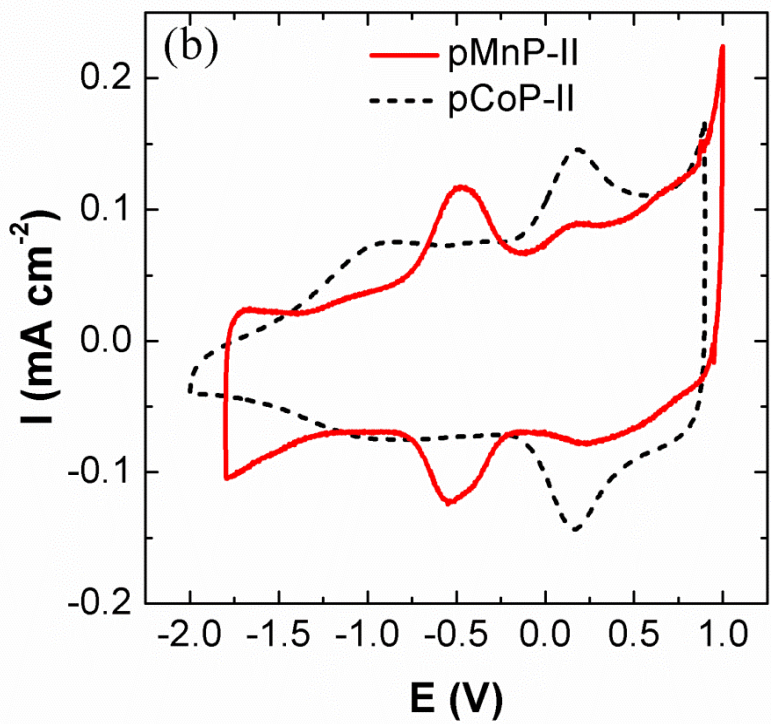

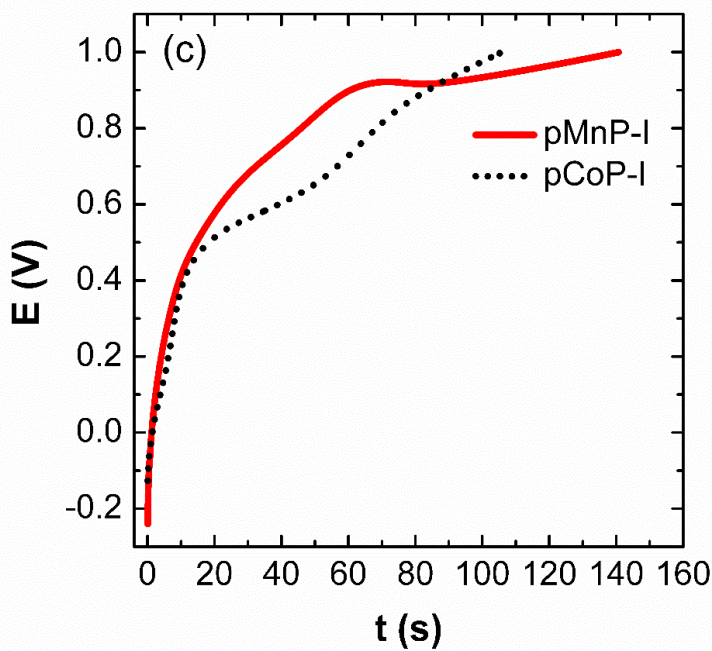

**Fig. 4.** Comparative cyclic voltammetric responses of Pt electrode coated with pMPs films (M = Mn or Co) of type I (deposition charge of pMgP-I: 5 mC cm$^{-2}$) (a) and type II (b) in 0.1 mol dm$^{-3}$ TBAPF$_6$ + AN solution. Scan rate: 0.1 V s$^{-1}$. (c) Typical potential-time dependence in the course of the galvanostatic transformation of pCoP-I and pMnP-I in 0.1 mol dm$^{-3}$ TBAPF$_6$ + AN solution. Applied anodic current density: 0.057 mA cm$^{-2}$ for both films.

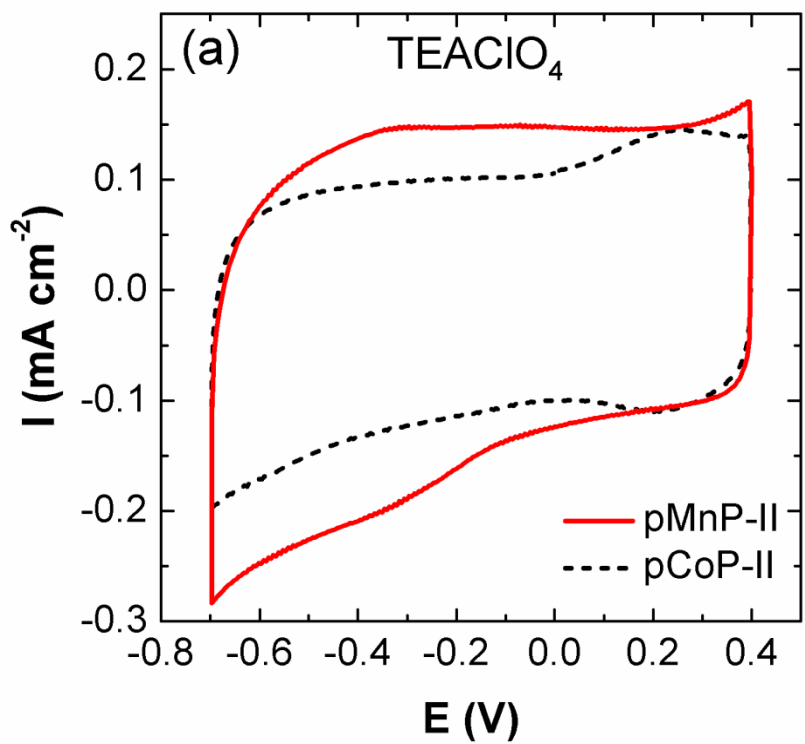

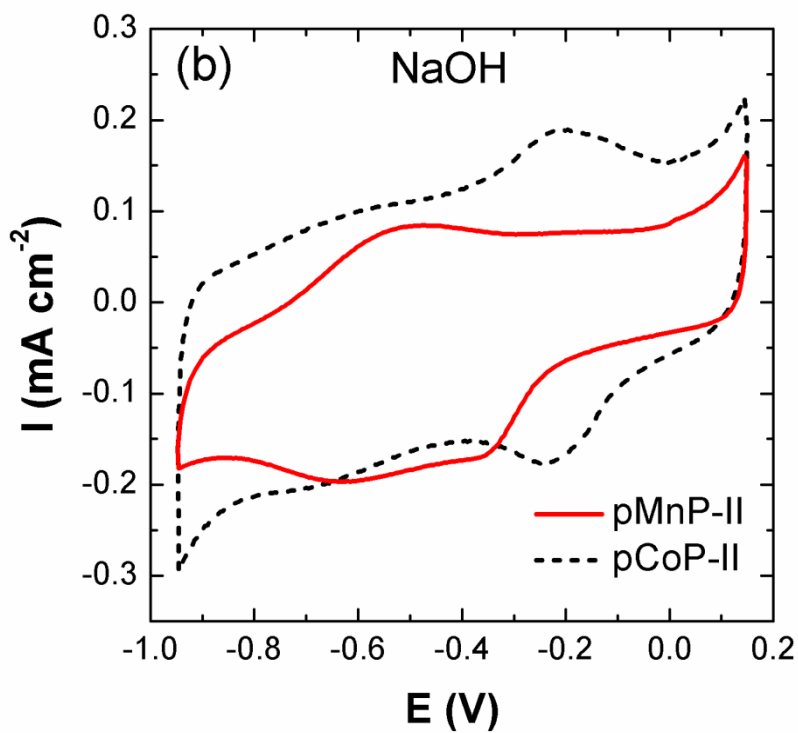

**Fig. 5.** Cyclic voltammetric curves of pMnP-II and pCoP-II coated glassy carbon electrodes (deposition charge of pMgP-I: 5 mC cm$^{-2}$) in deaerated aqueous solutions of 0.1 mol dm$^{-3}$ TEAClO$_4$ (a) and 0.1 mol dm$^{-3}$ NaOH (b). Scan rate: 0.1 V s$^{-1}$.

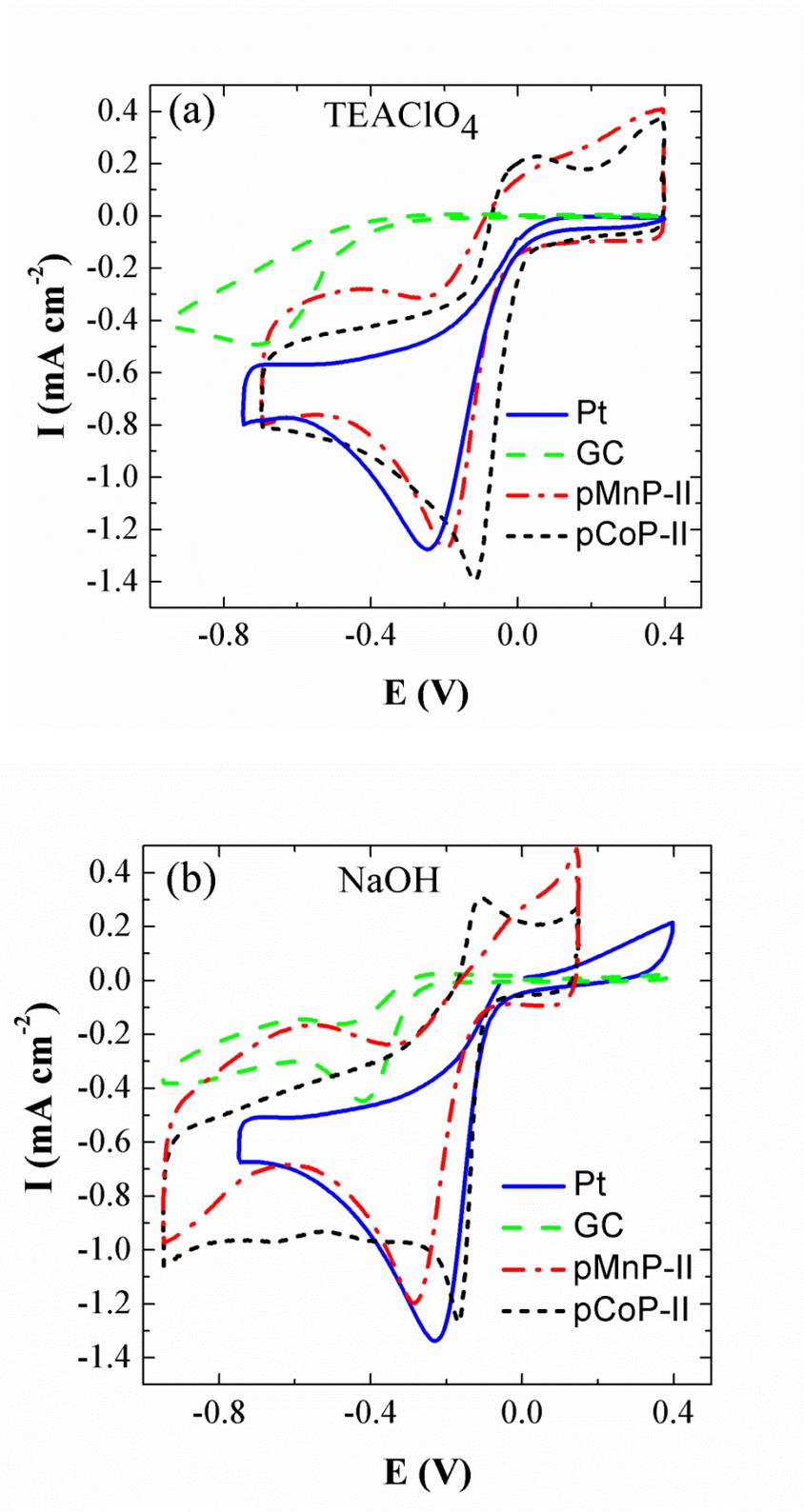

**Fig. 6.** Cyclic voltammograms (first cycle) of pMnP-II and pCoP-II films (deposition charge of pMgP-I: 5 mC cm$^{-2}$) on glassy carbon electrodes in oxygen-saturated aqueous electrolytes: (a) 0.1 mol dm$^{-3}$ TEAClO$_4$; (b) 0.1 mol dm$^{-3}$ NaOH. Electrochemical responses of bare platinum and glassy carbon electrodes of the same surface area are shown for comparison. Potential sweep rate: 0.1 V s$^{-1}$. Temperature: 25°C.

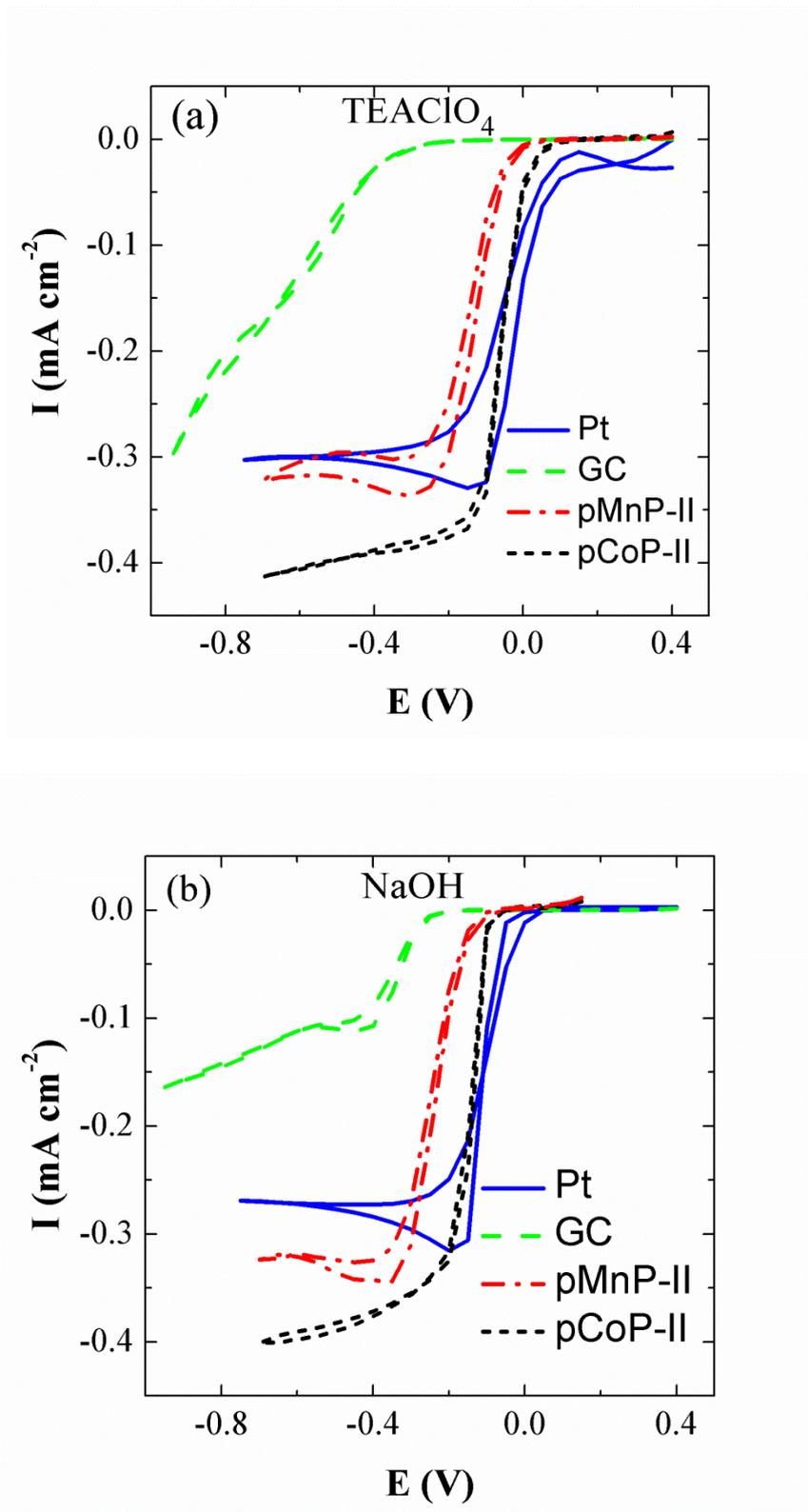

**Fig. 7.** Steady-state voltammograms of pMnP-II and pCoP-II films (deposition charge of pMgP-I: 5 mC cm$^{-2}$) on glassy carbon electrodes in oxygen-saturated aqueous electrolytes of (a) 0.1 mol dm$^{-3}$ TEAClO$_4$ and (b) 0.1 mol dm$^{-3}$ NaOH. Steady-state responses of bare platinum and glassy carbon electrodes of the same surface area are included for comparison. Temperature: 25°C. These measurements were performed

by means of a set of 50 mV potential steps where the potential value was kept constant for 30 s after the step while the "steady-state current" was determined as the average current during the last 2 s before the next potential step.

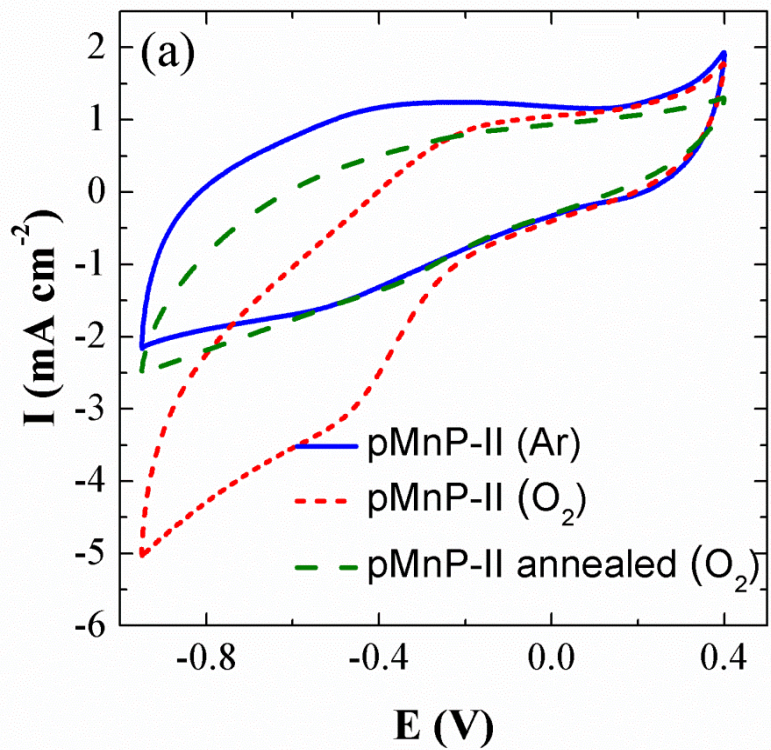

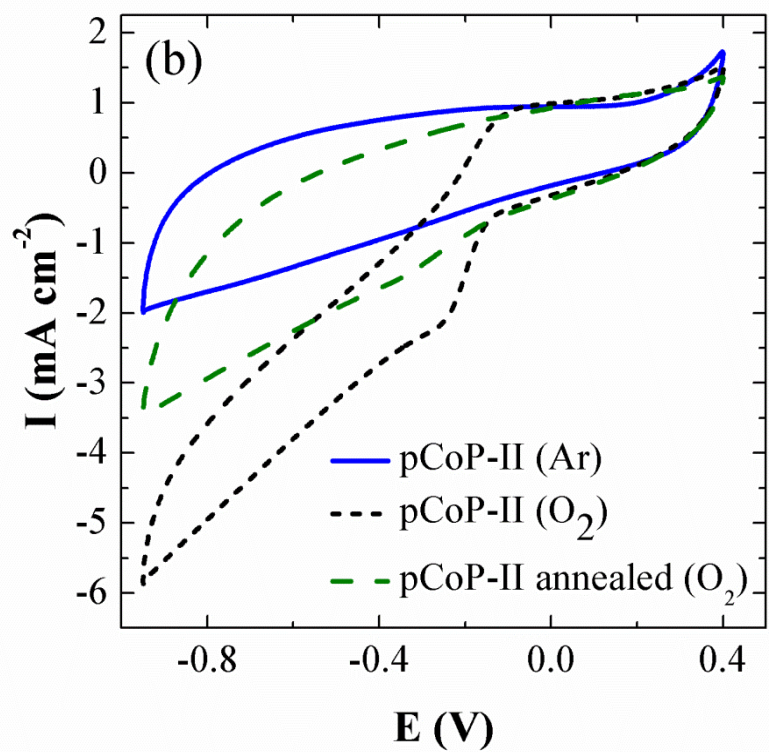

**Fig. 8.** Voltammetric curves of oxygen reduction (10$^{th}$ scan) at roughened glassy carbon electrodes covered with films of pMnP-II (a) and pCoP-II (b) recorded at 25 °C in alkaline electrolyte, before and after annealing treatment. Scan rate: 0.1 V s$^{-1}$.

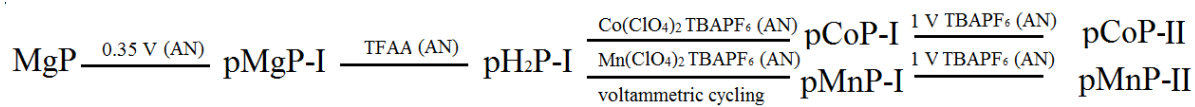

**Scheme 1.** Four-step procedure leading to formation of polyporphines of type II (pCoP-II and pMnP-II)

**Tables**

**Table 1.** Relative content of metal ions (M=Mg, Mn) in pMPs films derived from XPS measurements.

|  | at %, XPS | | | C/N | N/M |
|---|---|---|---|---|---|
|  | C | N | Me |  |  |
| $C_{20}H_{12}N_4M$(calculated) | - | - | - | 5.0 | 4.0 |
| pMgP-I | 85.1 | 11.6 | 3.3 | 7.3 | 3.5 |
| pCoP-I | 85.0 | 11.7 | 3.2 | 7.2 | 3.6 |
| pCoP-II | 84.3 | 13.4 | 2.3 | 6.3 | 5.8 |
| pMnP-I | 84.8 | 12.7 | 2.5 | 6.7 | 5.1 |
| pMnP-II | 83.5 | 13.5 | 3.1 | 6.2 | 4.4 |

**Table 2.** Maximum peak potentials ($E_p$) and half-wave potentials ($E_{1/2}$) values of oxygen reduction derived from Figs. 6 and 7.

|  | $E_p/E_{1/2}$, V | | | |
|---|---|---|---|---|
| Electrolyte | GC | Pt | pMnP-II | pCoP-II |
| TEAClO$_4$ | -0.70/-0.52 | -0.25/-0.02 | -0.20/-0.13 | -0.12/-0.05 |
| NaOH | -0.42/-0.33 | -0.23/-0.1 | -0.29/-0.23 | -0.17/-0.14 |